\documentstyle[12pt]{article}

\input amssym.tex

\begin{document}

\renewcommand{\a}{\alpha}
\renewcommand{\b}{\beta}
\newcommand{\g}{\gamma}           \newcommand{\G}{\Gamma}
\renewcommand{\d}{\delta}         \newcommand{\D}{\Delta}
\newcommand{\ve}{\varepsilon}
\newcommand{\eps}{\epsilon}
\newcommand{\k}{\kappa}
\newcommand{\ld}{\lambda}        \newcommand{\LD}{\Lambda}
\newcommand{\om}{\omega}         \newcommand{\OM}{\Omega}
\newcommand{\p}{\psi}             \newcommand{\PS}{\Psi}
\newcommand{\ro}{\rho}
\newcommand{\s}{\sigma}           \renewcommand{\S}{\Sigma}
\newcommand{\th}{\theta}         \newcommand{\T}{\Theta}
\newcommand{\f}{{\phi}}           \newcommand{\F}{{\Phi}}
\newcommand{\vf}{{\varphi}}
\newcommand{\y}{{\upsilon}}       \newcommand{\Y}{{\Upsilon}}
\newcommand{\z}{\zeta}
\newcommand{\X}{\Xi}
\newcommand{\cA}{{\cal A}}
\newcommand{\cB}{{\cal B}}
\newcommand{\cC}{{\cal C}}
\newcommand{\cD}{{\cal D}}
\newcommand{\cE}{{\cal E}}
\newcommand{\cF}{{\cal F}}
\newcommand{\cG}{{\cal G}}
\newcommand{\cH}{{\cal H}}
\newcommand{\cI}{{\cal I}}
\newcommand{\cJ}{{\cal J}}
\newcommand{\cK}{{\cal K}}
\newcommand{\cL}{{\cal L}}
\newcommand{\cM}{{\cal M}}
\newcommand{\cN}{{\cal N}}
\newcommand{\cO}{{\cal O}}
\newcommand{\cP}{{\cal P}}
\newcommand{\cQ}{{\cal Q}}
\newcommand{\cS}{{\cal S}}
\newcommand{\cR}{{\cal R}}
\newcommand{\cT}{{\cal T}}
\newcommand{\cU}{{\cal U}}
\newcommand{\cV}{{\cal V}}
\newcommand{\cW}{{\cal W}}
\newcommand{\cX}{{\cal X}}
\newcommand{\cY}{{\cal Y}}
\newcommand{\cZ}{{\cal Z}}
\newcommand{\hA}{{\widehat A}}
\newcommand{\hB}{{\widehat B}}
\newcommand{\hC}{{\widehat C}}
\newcommand{\hD}{{\widehat D}}
\newcommand{\hE}{{\widehat E}}
\newcommand{\hF}{{\widehat F}}
\newcommand{\hG}{{\widehat G}}
\newcommand{\hH}{{\widehat H}}
\newcommand{\hI}{{\widehat I}}
\newcommand{\hJ}{{\widehat J}}
\newcommand{\hK}{{\widehat K}}
\newcommand{\hL}{{\widehat L}}
\newcommand{\hM}{{\widehat M}}
\newcommand{\hN}{{\widehat N}}
\newcommand{\hO}{{\widehat O}}
\newcommand{\hP}{{\widehat P}}
\newcommand{\hQ}{{\widehat Q}}
\newcommand{\hS}{{\widehat S}}
\newcommand{\hR}{{\widehat R}}
\newcommand{\hT}{{\widehat T}}
\newcommand{\hU}{{\widehat U}}
\newcommand{\hV}{{\widehat V}}
\newcommand{\hW}{{\widehat W}}
\newcommand{\hX}{{\widehat X}}
\newcommand{\hY}{{\widehat Y}}
\newcommand{\hZ}{{\widehat Z}}
\newcommand{\Ha}{{\widehat a}}
\newcommand{\Hb}{{\widehat b}}
\newcommand{\Hc}{{\widehat c}}
\newcommand{\Hd}{{\widehat d}}
\newcommand{\He}{{\widehat e}}
\newcommand{\Hf}{{\widehat f}}
\newcommand{\Hg}{{\widehat g}}
\newcommand{\Hh}{{\widehat h}}
\newcommand{\Hi}{{\widehat i}}
\newcommand{\Hj}{{\widehat j}}
\newcommand{\Hk}{{\widehat k}}
\newcommand{\Hl}{{\widehat l}}
\newcommand{\Hm}{{\widehat m}}
\newcommand{\Hn}{{\widehat n}}
\newcommand{\Ho}{{\widehat o}}
\newcommand{\Hp}{{\widehat p}}
\newcommand{\Hq}{{\widehat q}}
\newcommand{\Hs}{{\widehat s}}
\newcommand{\Hr}{{\widehat r}}
\newcommand{\Ht}{{\widehat t}}
\newcommand{\Hu}{{\widehat u}}
\newcommand{\Hv}{{\widehat v}}
\newcommand{\Hw}{{\widehat w}}
\newcommand{\Hx}{{\widehat x}}
\newcommand{\Hy}{{\widehat y}}
\newcommand{\Hz}{{\widehat z}}
\newcommand{\deff}{\,\stackrel{\rm def}{\equiv}\,}
\newcommand{\lra}{\longrightarrow}
\newcommand{\ra}{\,\rightarrow\,}
\def\limar#1#2{\,\raise0.3ex\hbox{$\longrightarrow$\kern-1.5em\raise-1.1ex
\hbox{$\scriptstyle{#1\rightarrow #2}$}}\,}
\def\limarr#1#2{\,\raise0.3ex\hbox{$\longrightarrow$\kern-1.5em\raise-1.3ex
\hbox{$\scriptstyle{#1\rightarrow #2}$}}\,}
\def\limlar#1#2{\ \raise0.3ex
\hbox{$-\hspace{-0.5em}-\hspace{-0.5em}-\hspace{-0.5em}
\longrightarrow$\kern-2.7em\raise-1.1ex
\hbox{$\scriptstyle{#1\rightarrow #2}$}}\ \ }
\newcommand{\limm}[2]{\lim_{\stackrel{\scriptstyle #1}{\scriptstyle #2}}}
\newcommand{\wt}{\widetilde}
\newcommand{\os}{{\otimes}}
\newcommand{\da}{{\dagger}}
\newcommand{\stimes}{\times\hspace{-1.1 em}\supset}
\def\h{\hbar}
\newcommand{\ih}{\frac{\i}{\h}}
\newcommand{\exx}[1]{\exp\left\{ {#1}\right\}}
\newcommand{\ord}[1]{\mbox{\boldmath{$\cO$}}\left({#1}\right)}
\newcommand{\one}{{\leavevmode{\rm 1\mkern -5.4mu I}}}
\newcommand{\Z}{Z\!\!\!Z}
%
\newcommand{\Ibb}[1]{ {\rm I\ifmmode\mkern
            -3.6mu\else\kern -.2em\fi#1}}
\newcommand{\ibb}[1]{\leavevmode\hbox{\kern.3em\vrule
     height 1.2ex depth -.3ex width .2pt\kern-.3em\rm#1}}
\newcommand{\N}{{\Ibb N}}
\newcommand{\C}{{\ibb C}}
\newcommand{\R}{{\Ibb R}}
\newcommand{\HH}{{\Ibb H}}
\newcommand{\rational}{{\kern .1em {\raise .47ex
\hbox{$\scripscriptstyle |$}}
    \kern -.35em {\rm Q}}}
\newcommand{\bm}[1]{\mbox{\boldmath${#1}$}}
\newcommand{\intf}{\int_{-\infty}^{\infty}\,}
\newcommand{\LL}{\cL^2(\R^2)}
\newcommand{\LLS}{\cL^2(S)}
\newcommand{\Ree}{{\cal R}\!e \,}
\newcommand{\Imm}{{\cal I}\!m \,}
\newcommand{\tr}{{\rm {Tr} \,}}
\newcommand{\er}{{\rm{e}}}
\renewcommand{\i}{{\rm{i}}}
\newcommand{\divv}{{\rm {div} \,}}
\newcommand{\id}{{\rm{id}\,}}
\newcommand{\ad}{{\rm{ad}\,}}
\newcommand{\Ad}{{\rm{Ad}\,}}
\newcommand{\const}{{\rm{\,const\,}}}
\newcommand{\rank}{{\rm{\,rank\,}}}
\newcommand{\diag}{{\rm{\,diag\,}}}
\newcommand{\sign}{{\rm{\,sign\,}}}
\newcommand{\pa}{\partial}
\newcommand{\pad}[2]{{\frac{\partial #1}{\partial #2}}}
\newcommand{\padd}[2]{{\frac{\partial^2 #1}{\partial {#2}^2}}}
\newcommand{\paddd}[3]{{\frac{\partial^2 #1}{\partial {#2}\partial {#3}}}}
\newcommand{\der}[2]{{\frac{{\rm d} #1}{{\rm d} #2}}}
\newcommand{\derr}[2]{{\frac{{\rm d}^2 #1}{{\rm d} {#2}^2}}}
\newcommand{\fud}[2]{{\frac{\delta #1}{\delta #2}}}
\newcommand{\fudd}[2]{{\frac{\d^2 #1}{\d {#2}^2}}}
\newcommand{\fuddd}[3]{{\frac{\d^2 #1}{\d {#2}\d {#3}}}}
\newcommand{\dpad}[2]{{\displaystyle{\frac{\partial #1}{\partial #2}}}}
\newcommand{\dfud}[2]{{\displaystyle{\frac{\delta #1}{\delta #2}}}}
\newcommand{\dd}{\partial^{(\ve)}}
\newcommand{\ddd}{\bar{\partial}^{(\ve)}}
\newcommand{\dfrac}[2]{{\displaystyle{\frac{#1}{#2}}}}
\newcommand{\dsum}[2]{\displaystyle{\sum_{#1}^{#2}}}
\newcommand{\dint}{\displaystyle{\int}}
\newcommand{\dg}{\!\not\!\partial}
\newcommand{\vg}[1]{\!\not\!#1}
\def\<{\langle}
\def\>{\rangle}
\def\lgl{\langle\langle}
\def\rgr{\rangle\rangle}
\newcommand{\bra}[1]{\left\langle {#1}\right|}
\newcommand{\ket}[1]{\left| {#1}\right\rangle}
\newcommand{\vev}[1]{\left\langle {#1}\right\rangle}
\newcommand{\be}{\begin{equation}}
\newcommand{\ee}{\end{equation}}
\newcommand{\bn}{\begin{eqnarray}}
\newcommand{\en}{\end{eqnarray}}
\newcommand{\bnn}{\begin{eqnarray*}}
\newcommand{\enn}{\end{eqnarray*}}
\newcommand{\e}{\label}
\newcommand{\nbr}{\nonumber\\[2mm]}
\newcommand{\r}[1]{(\ref{#1})}
\newcommand{\refp}[1]{\ref{#1}, page~\pageref{#1}}
\renewcommand {\theequation}{\thesection.\arabic{equation}}
\renewcommand {\thefootnote}{\fnsymbol{footnote}}
\newcommand{\qq}{\qquad}
\newcommand{\qqq}{\quad\quad}
\newcommand{\biz}{\begin{itemize}}
\newcommand{\eiz}{\end{itemize}}
\newcommand{\ben}{\begin{enumerate}}
\newcommand{\een}{\end{enumerate}}
\def\nc{noncommutative }
\def\ncy{noncommutativity }
\def\com{commutative }
\def\JLD{Jost-Lehmann-Dyson }
\def\th{$\theta_{\mu\nu}$}
\def \simlt{\stackrel{<}{{}_\sim}}
\def\ss{$\theta_{0i}=0$}
\def\P{Poincar\'e}
\thispagestyle{empty}
\begin{flushright}
\end{flushright}

\begin{center}

{\Large{\bf{Jost-Lehmann-Dyson Representation\\
and Froissart-Martin Bound\\
in Quantum Field Theory\\ on Noncommutative Space-Time
\footnote{Work dedicated to the memory of Rolf Hagedorn; based on
the talk given by M. Chaichian in the "Hagedorn Memorial Meeting",
CERN, 28 November 2003,
http://wwwth.cern.ch/hagedorn/Hagedorn.htm}}}} \vskip .7cm

{\bf{\large{M. Chaichian and A. Tureanu}}

{\it High Energy Physics Division, Department of Physical
Sciences,
University of Helsinki\\
\ \ {and}\\
\ \ Helsinki Institute of Physics, P.O. Box 64, FIN-00014
Helsinki, Finland}}

\end{center}

\setcounter{footnote}{0}

{\bf Abstract} In the framework of quantum field theory (QFT) on
noncommutative (NC) space-time with $SO(1,1)\times SO(2)$
symmetry, which is the feature arising when one has only
space-space noncommutativity ($\theta_{0i}=0$), we prove that the
Jost-Lehmann-Dyson representation, based on the causality
condition usually taken in connection with this symmetry, leads to
the mere impossibility of drawing any conclusion on the
analyticity of the $2\rightarrow 2$-scattering amplitude in
$\cos\Theta$, $\Theta$ being the scattering angle. A physical
choice of the causality condition rescues the situation and as a
result an analog of Lehmann's ellipse as domain of analyticity in
$\cos\Theta$ is obtained. However, the enlargement of this
analyticity domain to Martin's ellipse and the derivation of the
Froissart bound for the total cross-section in NC QFT is possible
{\it only} in the special case when the incoming momentum is
orthogonal to the NC plane. This is the first example of a
nonlocal theory in which the cross-sections are subject to a
high-energy bound. For the general configuration of the direction
of the incoming particle, although the scattering amplitude is
still analytic in the Lehmann ellipse, no bound on the total
cross-section has been derived. This is due to the lack of a
simple unitarity constraint on the partial-wave amplitudes, which
could be used in this case. High-energy upper bounds on the total
cross-section, among others, are also obtained for an arbitrary
flat (noncompact) dimension of NC space-time.


\section{Introduction}

The development of QFT on NC space-time, especially after the
seminal work of Seiberg and Witten \cite{SW}, which showed that
the NC QFT arises from string theory, has triggered lately the
interest also towards the formulation of an axiomatic approach to
the subject. Consequently, the analytical properties of scattering
amplitude in energy $E$ and forward dispersion relations have been
considered \cite{Liao, CMTV}, Wightman functions have been
introduced and the CPT theorem has been proven \cite{AG, CPT}, and
as well attempts towards a proof of the spin-statistics theorem
have been made \cite{CPT}\footnote {In the context of the
Lagrangian approach to NC QFT, the CPT and spin-statistics
theorems have been proven in general in \cite{CNT}; for CPT
invariance in NC QED, see \cite{Shahin,Kostelecky}, and in NC
Standard Model \cite{Aschieri}.}.

In the axiomatic approach to commutative QFT, one of the
fundamental results consisted of the rigorous proof of the
Froissart bound on the high-energy behaviour of the scattering
amplitude, based on its analyticity properties \cite{Froissart,
Martin}. In this paper we aim at  obtaining the analog of this
bound when the space-time is noncommutative. Such an achievement,
besides being topical in itself, will also prove fruitful in the
conceptual understanding of subtle issues, such as causality, in
nonlocal theories to which the NC QFT's belong.

In the following we shall consider NC QFT on a space-time with the
commutation relation
\be\label{cr}[x_\mu,x_\nu]=i\theta_{\mu\nu}\ , \ee
where $\theta_{\mu\nu}$ is an antisymmetric constant matrix (for a
review, see, e.g., \cite{DN,Szabo}). Such NC theories violate
Lorentz invariance, while translational invariance still holds. We
can always choose the system of coordinates, such that
$\theta_{13}=\theta_{23}=0$ and
$\theta_{12}=-\theta_{21}\equiv\theta$. Then, for the particular
case of space-space noncommutativity, i.e. $\theta_{0i}=0$, the
theory is invariant under the subgroup $SO(1,1)\times SO(2)$ of
the Lorentz group. The requirement that time be commutative
($\theta_{0i}=0$) discards the well-known problems with the
unitarity \cite{unit} of the NC theories and with causality
\cite{causal1,causal2}. As well, the $\theta_{0i}=0$ case allows a
proper definition of the $S$-matrix \cite{CMTV}.

In the conventional (commutative) QFT, the Froissart bound was
first obtained \cite{Froissart} using the conjectured Mandelstam
representation (double dispersion relation) \cite{mandel}, which
assumes analyticity in the entire $E$ and $\cos\Theta$ complex
planes. The Froissart bound,
\be\label{fm} \sigma_{tot}(E)\leq c\ \ \ln^2 \frac{E}{E_0}\ , \ee
expresses the upper limit of the total cross-section
$\sigma_{tot}$ as a function of the CMS energy $E$, when
$E\to\infty$. However, such an analyticity or equivalently the
double dispersion relation has not been proven, while smaller
domains of analyticity in $\cos\Theta$ were already known
\cite{Lehmann}.

One of the main ingredients in rigorously obtaining the Froissart
bound is the Jost-Lehmann-Dyson representation \cite{JL,D} of the
Fourier transform of the matrix element of the commutator of
currents, which is based on the causality as well as the spectral
conditions (for an overall review, see \cite{Schweber}). Based on
this integral representation, one obtains the domain of
analyticity of the scattering amplitude in $\cos\Theta$. This
domain proves to be an ellipse $-$ the so-called Lehmann's ellipse
\cite{Lehmann}. Using this domain of analyticity, Greenberg and
Low \cite{GL} found a weaker bound on the high-energy behaviour of
the total cross-section:
\be\label{gl} \sigma_{tot}(E)\leq c\ \ E^2 \ln^2 \frac{E}{E_0}\ .
\ee
However, the domain of analyticity in $\cos\Theta$ can be enlarged
to the so-called Martin's ellipse by using the dispersion
relations satisfied by the scattering amplitude and the unitarity
constraint on the partial-wave amplitudes. Using this larger
domain of analyticity, the Froissart bound (\ref{fm}) was
rigorously proven \cite{Martin} (for a review, see \cite{review}).

Further on, the analog of the Froissart-Martin bound was
rigorously obtained for the $2 \to 2$-particle scattering in a
space-time of arbitrary dimension $D$ \cite{Fischer,CFV}.

In NC QFT with $\theta_{0i}=0$ we shall follow the same path for
the derivation of the high-energy bound on the scattering
amplitude, starting from the Jost-Lehmann-Dyson representation and
adapting the derivation to the new symmetry $SO(1,1)\times SO(2)$
and to the nonlocality of the NC theory. In Section 2 we derive
the Jost-Lehmann-Dyson representation satisfying the light-wedge
(instead of light-cone) causality condition, which has been used
so far, being inspired by the above symmetry. In Section 3 we show
that no analyticity of the scattering amplitude in $\cos\Theta$
can be obtained in such a case. However, with a newly introduced
causality condition, based on physical arguments, we obtain from
the Jost-Lehmann-Dyson representation a domain of analyticity in
$\cos\Theta$, which coincides with the Lehmann ellipse. In Section
4 we show that the extension of this analyticity domain to
Martin's ellipse is possible in the case of the incoming particle
momentum orthogonal to the NC plane $(x_1,x_2)$, which eventually
enables us to derive rigorously the analog of the Froissart-Martin
bound (\ref{fm}) for the total cross-section. The general
configuration of incoming particle momentum is also discussed,
together with the problems which arise. In Section 5 we consider
higher-dimensional NC theories and obtain the high-energy bounds
for a particular setting, in the case of even dimension D; we also
discuss the peculiarities of the NC theories on space-time of
higher odd dimension. Section 6 is devoted to conclusion and
discussions.

\setcounter{equation}{0}

\section{Jost-Lehmann-Dyson representation}

The Jost-Lehmann-Dyson representation \cite{JL,D} is the integral
representation for the Fourier transform of the matrix element of
the commutator of currents:
\be\label{fq} f(q)=\int d^4xe^{iqx}f(x)\ , \ee
where
\be\label{fx} f(x)=\langle
p'|[j_1(\frac{x}{2}),j_2(-\frac{x}{2})]|p\rangle\ ,\ee
satisfying the causality and spectral conditions. The process
considered is the $2\to 2$  scalar particles scattering, $k+p\to
k'+p'$, and $j_1$ and $j_2$ are the scalar currents corresponding
to the incoming and outgoing particles with momenta $k$ and $k'$
(see also \cite {Schweber,BS}).

For NC QFT with $SO(1,1)\times SO(2)$ symmetry, in \cite{LAG} a
new causality condition was proposed, involving (instead of the
light-cone) the light-wedge corresponding to the coordinates $x_0$
and $x_3$, which form a two-dimensional space with the $SO(1,1)$
symmetry. Accordingly we shall require the vanishing of the
commutator of two currents (in general, observables) at space-like
separations in the sense of $SO(1,1)$ as:
\be\label{causal} [j_1(\frac{x}{2}),j_2(-\frac{x}{2})]=0 \ ,\ \
\mbox{for}\ \ \tilde x^2\equiv x_0^2-x_3^2<0\ . \ee

The spectral condition compatible with (\ref{causal}) would
require now that the physical momenta be in the forward
light-wedge:
\be\label{spectr} \tilde p^2\equiv p_0^2-p_3^2>0\ \ \mbox{and}\ \
p_0>0\ . \ee

The spectral condition (\ref{spectr}) will impose restrictions on
$f(q)$. Using the translational invariance in (\ref{fx}), one can
express the matrix element of the commutator of currents, $f(x)$,
in the form:
\bn\label{tr_inv} f(x)&=&\int dq
e^{-iqx+i(p+p')\frac{x}{2}}G_1(q)-\int dq
e^{iqx-i(p+p')\frac{x}{2}}G_2(q)\cr &=&\int dq
e^{-iqx}\left[G_1\left(q+\frac{1}{2}(p+p')\right)-G_2\left(-q+\frac{1}{2}(p+p')\right)\right]\
, \en
where
\bn\label{gs}
 G_1(q)&=&\langle p'|j_1(0)|q\rangle\langle
q|j_2(0)|p\rangle\ ,\cr G_2(q)&=&\langle p'|j_2(0)|q\rangle\langle
q|j_1(0)|p\rangle\ . \en
Comparing (\ref{tr_inv}) with the inverse Fourier
transformation\footnote{Throughout the paper we omit all the
inessential factors of $(2\pi)^n$, which are irrelevant for the
analyticity considerations.}, $f(x)=\int dq e^{-iqx}f(q)$, it
follows that
\be f(q)=f_1(q)-f_2(q)=
G_1\left(q+\frac{1}{2}(p+p')\right)-G_2\left(-q+\frac{1}{2}(p+p')\right)\
.\ee
Given the way the functions $G_1$ and $G_2$ are defined in
(\ref{gs}), one finds that  $f(q)=0$ in the region where the
momenta $q+\frac{1}{2}(p+p')$ and $-q+\frac{1}{2}(p+p')$ are
simultaneously nonphysical, i.e. when they are out of the future
light-wedge (\ref{spectr}).

In order to express the condition for $f(q)=0$, we shall define
the $SO(1,1)$-invariant $\tilde m^2=k^2_0-k^2_3=f(m^2,
k_1^2+k_2^2)$, where $k$ is the momentum of an arbitrary state and
$m$ is its mass. However, we have to point out that $\tilde m$ is
only a kinematical variable, invariant with respect to $SO(1,1)$
(but not the mass).

For the {\it physical} states with momentum $q+\frac{1}{2}(p+p')$,
we take $\tilde m_{1}$ to be the minimal value of the
$SO(1,1)$-invariant quantity above. Then, in the Breit frame,
where $\frac{1}{2}(p+p')=(p_0,0,0,0)$, one finds that $f_1(q)\neq
0$ for all the $q$ values, satisfying the spectral condition
$q_0+p_0\geq 0$ and $(q_0-p_0)^2-q_3^2\geq 0$. In other words,
$f_1(q)=0$ for $q_0<-p_0+\sqrt{q_3^2+\tilde m_1^2}$. Similarly one
finds that $f_2(q)=0$ for $p_0-\sqrt{q_3^2+\tilde m_2^2}<q_0$
(where $\tilde m_2$ has a meaning analogous to that of $\tilde
m_1$, but for the states with the momentum
$-q+\frac{1}{2}(p+p')$).

As a result, due to the spectral condition (\ref{spectr}),
$f(q)=0$ in the region outside the hyperbola
\be\label{4.11'} p_0-\sqrt{q_3^2+\tilde
m_2^2}<q_0<-p_0+\sqrt{q_3^2+\tilde m_1^2}\ . \ee

To derive the \JLD representation, further we consider the
6-dimensional space-time with the Minkowskian metric
$(+,-,-,-,-,-)$. On this space, we define the vector
$z=(x_0,x_1,x_2,x_3,y_1,y_2)$. For practical purposes we introduce
also the notations for the 2-dimensional vector $\tilde x =
(x_0,x_3)$ and the 4-dimensional vector $\tilde z=(z_0, z_3,
z_4,z_5)\equiv(x_0, x_3, y_1,y_2)$. On the 6-dimensional space we
define the function
\be\label{F} F(z)=f(x)\delta(\tilde x^2-y^2)=f(x)\delta(\tilde
z^2), \ee
depending on all 6 coordinates.

When the causality condition (\ref{causal}) is fulfilled, i.e. for
the physical region, $f(x)$ and $F(z)$ determine each other, since
\begin{equation}\label{int}
\int dy_1dy_2 F(z)=f(x)\theta(\tilde x^2) =\left
\{\begin{tabular}{lll}
$f(x)$ & for & $\tilde x^2>0\ ,$ \\
0 &
for & $\tilde x^2<0\ .$\\
\end{tabular}
\right.
\end{equation}

The Fourier transform of $F(z)$,
\be F(r)=\int d^6ze^{izr}F(z)\ , \ee
can be expressed, using (\ref{F}) and (\ref{int}), as
\be\label{4.14} F(r)=\int d^4qD_1(r-\hat q)f(q)\ . \ee
Denoting the remaining 4-dimensional vector $\tilde r=(r_0, r_3,
r_4, r_5)$, we have
\be\label{D_1} D_1(r)=\int d^6ze^{izr}\delta(\tilde
z^2)=\frac{\delta(r_1)\delta(r_2)}{\tilde r^2}
=\delta(r_1)\delta(r_2)D_1(\tilde r)\ , \ee
with $D_1(\tilde r)=\frac{1}{\tilde r^2}$.

We define now the "subvector" of a 6-dimensional vector as $\hat
q=(q_0,q_1,q_2,q_3,0,0)$ and we find the relation between $F(\hat
q)$ and $f(q)$ in view of the causality condition (\ref{causal}):
\be\label{4.14'} F(\hat q)=\int d^4x f(x)\theta(\tilde
x^2)e^{iqx}=f(q)\ . \ee
$D_1(\tilde r)$ satisfies the 4-dimensional wave-equation:
\be \Box _4 D_1(\tilde r)=0\ , \ee
where the d'Alembertian is defined with respect to the coordinates
$r_0, r_3, r_4, r_5$. Then, due to (\ref{4.14}), it follows that
$F(r)$ satisfies the same equation,
\be\label{difeq} \Box _4F(r)=0\ . \ee
It is crucial to note that $F(r)$ depends on all 6 variables $r_0,
...r_5$:
$$
F(r)=\int d^4q f(q)D_1(\tilde r-\tilde
q)\delta(r_1-q_1)\delta(r_2-q_2)\ ,
$$
where $\tilde q=(q_0, q_3, 0, 0)$.

The solution of (\ref{difeq}) can be written in the form
\cite{Vlad}:
$$
F(r')=\int d^3\Sigma_\alpha\int\int d r_1
dr_2\left[F(r)\frac{\partial D(\tilde r-\tilde r')}{\partial
\tilde r_\alpha} -D(\tilde r-\tilde r')\frac{\partial
F(r)}{\partial \tilde r_\alpha}\right]\delta(r_1)\delta(r_2)\ ,
$$
where $D(\tilde r)$ satisfies the homogeneous differential
equation $\Box _4D(\tilde r)=0$, with the initial conditions
$$
D(\tilde r)|_{r_0=0}=0 \ \ \ \mbox{and}\ \ \ \frac{\partial
D}{\partial r_0}(\tilde r)|_{r_0=0}=\prod_{i=1}^3\delta(r_i)\ .
$$
The first condition implies that $D(\tilde r)$ is an odd function,
with the result that:
\be\label{Dr} D(\tilde r)=\int d^4ze^{-i\tilde z\tilde
r}\epsilon(z_0)\delta(\tilde z^2)=\epsilon(r_0)\delta(\tilde r^2).
\ee
We note here that the surface $\Sigma$ is 3-dimensional and not
5-dimensional as it is in the commutative case with light-cone
causality condition. Now we can express $f(q)$ using (\ref{4.14'})
as:
\bn f(q)=F(\hat q)=\int d r_1 d r_2 \delta
(r_1-q_1)\delta(r_2-q_2)\cr \times\int d^3\Sigma_\alpha
[F(r)\frac{\partial D(\tilde r-\tilde q)}{\partial \tilde
r_\alpha} -D(\tilde r-\tilde q)\frac{\partial F(r)}{\partial
\tilde r_\alpha}]\ . \en
Due to the arbitrariness of the surface $\Sigma$, one can reduce
the integration over $r_4$ and $r_5$, using the cylindrical
symmetry, to the integral over $\kappa^2=r_4^2+r_5^2$.
Subsequently we change the notation of variables $r_i$ to $u_i$
and use the explicit form of $D(\tilde r)$ from (\ref{Dr}) to
obtain:
\begin{eqnarray}\label{interm}
&&f(q)=\int d u_1 d u_2\delta(u_1-q_1)\delta(u_2-q_2)\int
d^1\Sigma_j d\kappa^2\cr
&&\times\{F(u,\kappa^2)\frac{\partial}{\partial \tilde
u_j}\left[\epsilon(u_0-q_0)\delta((\tilde u-\tilde
q)^2-\kappa^2)\right]\cr &&-\epsilon(u_0-q_0)\delta((\tilde
u-\tilde q)^2-\kappa^2)\frac{\partial
F(u,\kappa^2)}{\partial\tilde u_j}\}\ .
\end{eqnarray}

Using the standard mathematical procedure \cite{Vlad} for
performing the integration in (\ref{interm}), we obtain the \JLD
representation in NC QFT, satisfying the light-wedge causality
condition (\ref{causal}):
\bn\label{4.18} f(q)&=&\int d^4u d\kappa^2\epsilon
(q_0-u_0)\delta[(q_0-u_0)^2-(q_3-u_3)^2-\kappa^2]\cr
&\times&\delta(q_1-u_1)\delta(q_2-u_2) \phi(u,\kappa^2)\ , \en
where $\phi(u,\kappa^2)=-\frac{\partial
F(u,\kappa^2)}{\partial\tilde u_0}$.

Equivalently, denoting $\tilde u=(u_0,u_3)$, (\ref{4.18}) can be
written as:

\be\label{4.18'} f(q)=\int d^2\tilde u d\kappa^2\epsilon
(q_0-u_0)\delta[(\tilde q-\tilde u)^2-\kappa^2] \phi(\tilde
u,q_1,q_2, \kappa^2)\ . \ee

The function $\phi(\tilde u, q_1,q_2, \kappa^2)$ is an arbitrary
function, except that the requirement of spectral condition
determines a domain in which $\phi(\tilde u,q_1,q_2, \kappa^2)=0$.
This domain is outside the region where the $\delta$ function in
(\ref{4.18'}) vanishes, i.e.
\be\label{4.19} (\tilde q-\tilde u)^2-\kappa^2=0\ , \ee
but with $\tilde q$ in the region given by (\ref{4.11'}), where
$f(q)=0$. Putting together (\ref{4.19}) and (\ref{4.11'}), we
obtain the domain out of which $\phi(\tilde u,q_1,q_2,
\kappa^2)=0$:
\begin{eqnarray}
&&a)\ \frac{1}{2}(\tilde p+\tilde p')\pm \tilde u\  \  \mbox{are in the forward light-wedge (cf. (\ref{spectr}))};\\
&&b)\ \kappa\geq max\left\{0, \tilde m_1-\sqrt{\left(\frac{\tilde
p+\tilde p'}{2}+\tilde u\right)^2}, \tilde
m_2-\sqrt{\left(\frac{\tilde p+\tilde p'}{2}-\tilde
u\right)^2}\right\}\ .\nonumber
\end{eqnarray}

For the purpose of expressing the scattering amplitude, we
actually need the Fourier transform $f_R(q)$ of the retarded
commutator,
\be\label{ret} f_R(x)=\theta(x_0)f(x)= \langle
p'|\theta(x_0)[j_1(\frac{x}{2}),j_2(-\frac{x}{2})]|p\rangle\ . \ee
Using  (\ref{ret}) and the Fourier transformation $f(x)=\int dq'
e^{-iq'x}f(q')$, we can express $f_R(q)$ as follows:
\bn\label{ret_q} f_R(q)&=&\int dx e^{iqx}f_R(x)=\int dx
e^{iqx}\theta(x_0)f(x)\cr &=& \int dq' f(q')\int dx
e^{i(q-q')x}\theta(x_0)\ . \en
Taking into account that
$$
\int dx_0 e^{i(q-q')x}\theta(x_0)=-i\frac{e^{i(\vec q-\vec q')\vec
x}}{q_0-q'_0}\ ,
$$
eq. (\ref{ret_q}) becomes:
$$ f_R(q)=i\int d q'_0 \frac{f(q'_0,\vec
q)}{q'_0-q_0}\ .$$

Now in the above formula we introduce the Jost-Lehmann-Dyson
representation (\ref{4.18'}), with the result:
\be\label{f_ret} f_R(q)=i\int \frac{d q'_0} {q'_0-q_0}\int
d^2\tilde u d\kappa^2\epsilon
(q'_0-u_0)\delta[(q'_0-u_0)^2-(q_3-u_3)-\kappa^2] \phi(\tilde
u,q_1,q_2, \kappa^2)\ .\ee
In (\ref{f_ret}) one can integrate over $q'_0$, using the known
formula of integration with a $\delta$-function, $\int
G(x)\delta(g(x))dx=\sum_i\frac{G(x_{0i})}{\frac{\partial
g}{\partial x}|_{x=x_{0i}}}$, where $x_{0i}$ are the simple roots
of the function $g(x)$. We identify in (\ref{f_ret})
$G(q'_0)=\frac{\epsilon(q'0-u_0)}{q'_0-q_0}$ and
$g(q'_0)=(q'_0-u_0)^2-(q_3-u_3)-\kappa^2$ (with the roots
$q'_0=u_0\pm\left[(q_3-u_3)^2+\kappa^2\right]^{1/2}$).

With these considerations, from (\ref{f_ret}) we obtain the NC
version of the \JLD representation for the retarded commutator:
\be\label{4.32} f_R(q)=\int d^2 \tilde u
d\kappa^2\frac{\phi(\tilde
u,q_1,q_2,\kappa^2)}{(q_0-u_0)^2-(q_3-u_3)^2-\kappa^2}\ . \ee
Compared to the usual \JLD representation,
\be\label{c_jld} f_R^{comm}(q)=\int d^4 u d\kappa^2\frac{\phi(
u,\kappa^2)}{(q_0-u_0)^2-(\vec q-\vec u)^2-\kappa^2}\ , \ee
the expression (\ref{4.32}) is essentially different in the sense
that the arbitrary function $\phi$ now depends on $q_1$ and $q_2$.
This feature will have further crucial implications in the
discussion of analyticity of the scattering amplitude in
$\cos\Theta$.

\section{ Analyticity of the scattering amplitude in
$\cos\Theta$. \\Lehmann's ellipse}

In the center-of-mass system (CMS) and in a set in which the
incoming particles are along the vector $\vec\beta=
(0,0,\theta)$\footnote{The 'magnetic' vector $\vec\beta$ is
defined as $\beta_i=\frac{1}{2}\epsilon_{ijk}\theta_{jk}$. The
terminology stems from the antisymmetric background field
$B_{\mu\nu}$ (analogous to $F_{\mu\nu}$ in QED), which gives rise
to noncommutativity in string theory, with $\theta_{\mu\nu}$
essentially proportional to $B_{\mu\nu}$ (see, e.g., \cite{SW}).},
the scattering amplitude in NC QFT depends still on only two
variables, the CM energy $E$ and the cosine of the scattering
angle, $\cos\Theta$ (for a discussion about the number of
variables in the scattering amplitude for a general type of
noncommutativity, see \cite{CMT}).

In terms of the \JLD representation, the scattering amplitude is
written as (cf. \cite{Schweber} for commutative case):
\be\label{scat} M(E,\cos\Theta)=i\int d^2\tilde u
d\kappa^2\frac{\phi(\tilde u, \kappa^2, k+p, (k'-p')_{1,2})}
{\left[\frac{1}{2}(\tilde k'-\tilde p')+\tilde
u\right]^2-\kappa^2}\ , \ee
where $\phi(\tilde u, \kappa^2,...)$ is a function of its
$SO(1,1)$- and $SO(2)$-invariant variables: $u_0^2-u_3^2$,
$(k_0+p_0)^2-(k_3-p_3)^2$, $(k_1+p_1)^2+(k_2+p_2)^2$,
$(k'_1-p'_1)^2+(k'_2-p'_2)^2$,... The function $\phi$ is zero in a
certain domain, determined by the causal and spectral conditions,
but otherwise arbitrary.

For the discussion of analyticity of $M(E, \cos\Theta)$ in
$\cos\Theta$, it is of crucial importance that all dependence on
$\cos\Theta$ be contained in the denominator of (\ref{scat}). But,
since the {\it arbitrary} function $\phi$ depends now on
$(k'-p')_{1,2}$, it also depends on $\cos\Theta$. This makes
impossible the mere consideration of any analyticity property of
the scattering amplitude in $\cos\Theta$.

One might wonder now whether in the above derivation there is any
condition which could be subject to challenge. In that case there
might also appear the possibility that an analyticity domain can
be obtained, which might lead to a high-energy upper bound on the
scattering amplitude, analogous to the Froissart-Martin bound. In
this connection, we would like to mention that all perturbative
calculations performed in NC QFT show that the scattering
amplitude respects the Froissart bound when $\theta_{0i}=0$ (see,
e.g., \cite{CMT}).

Since the \JLD representation reflects the effect of the causal
and spectral axioms, we notice that the hypotheses (\ref{causal})
and (\ref{spectr}) used for the present derivation of JLD
representation are too weak, in the sense of their physical
implications, since they allow for a much larger physical region,
by not at all taking into account the effect of the NC coordinates
$x_1$ and $x_2$. This remark is also in accord with the result of
\cite{Liao}, in which it was shown that the forward dispersion
relation cannot be obtained in NC QFT with the causality condition
(\ref{causal}), except for the case of incoming particle momentum
being orthogonal to the NC plane, i.e. in the direction of the
$\vec\beta$-vector.

\subsection{Causality in NC QFT}

In the following, we shall challenge the causality condition
(\ref{causal})
\be\label{caus} [j_1(\frac{x}{2}),j_2(-\frac{x}{2})]=0\ , \,\ \
\mbox{for}\ \  \tilde x^2\equiv x_0^2-x_3^2<0\ , \ee
which takes into account {\it only} the variables connected with
the $SO(1,1)$ symmetry.

This causality condition would be suitable in the case when
nonlocality in NC variables $x_1$ and $x_2$ is {\it infinite},
which is not the case on a space with the commutation relation
$[x_1,x_2]=i\theta$, which implies $\Delta x_1\Delta
x_2\geq\frac{\theta}{2}$. The fact that in the causality condition
(\ref{caus}) the coordinates $x_1$ and $x_2$ do not enter means
that the propagation of a signal in this plane is instantaneous:
{\it no matter how far apart two events are in the \nc
coordinates}, the allowed region for correlation is given by only
the condition $x_0^2-x_3^2>0$, which involves the propagation of a
signal only in the $x_3$-direction, while the time for the
propagation along $x_1$- and $x_2$-directions is totally ignored.
%

The uncertainty relation $\Delta x_1\Delta x_2\sim\theta$, which
follows from the considered commutation relation of the coordinates,
puts a lower bound on localization. Admitting that the scale of
nonlocality in $x_1$ and $x_2$ is $l\sim\sqrt\theta$, then the
propagation of interaction in the \nc coordinates is instantaneous
{\it only within this distance} $l$. It follows then that two events
are correlated, i.e. $[j_1(\frac{x}{2}),j_2(-\frac{x}{2})]\neq 0$,
when $x_1^2+x_2^2\leq l^2$ (where $x_1^2+x_2^2$ is the distance in
the NC plane with $SO(2)$ symmetry), provided also that
$x^2_0-x^2_3\geq0$ (the events are time-like separated in the sense
of $SO(1,1)$). Adding the two conditions, we obtain that
\be\label{nonloc} [j_1(\frac{x}{2}),j_2(-\frac{x}{2})]\neq 0\ , \
\mbox{for}\ \ x_0^2-x_3^2-(x_1^2+x_2^2-l^2)\geq 0\ .\ee
The negation of condition (\ref{nonloc}) leads to the conclusion
that the locality condition should indeed be given by:
$$
[j_1(\frac{x}{2}),j_2(-\frac{x}{2})]=0\ ,\ \ \ \mbox{for}\ \ \
\tilde x^2-(x_1^2+x_2^2-l^2)\equiv x_0^2-x_3^2-(x_1^2+x_2^2-l^2)<0\
,
$$
or, equivalently,
\be\label{newcaus} [j_1(\frac{x}{2}),j_2(-\frac{x}{2})]=0\ ,\
\mbox{for}\ x_0^2-x_3^2-(x_1^2+x_2^2)<-l^2\ , \ee
where $l^2$ is a constant proportional to NC parameter $\theta$.
When $l^2\rightarrow 0$, (\ref{newcaus}) becomes the usual
locality condition.

When $x_1^2+x_2^2>l^2$, for the propagation of a signal only the
difference $x_1^2+x_2^2-l^2$ is time-consuming and thus in the
locality condition it is the quantity
$x_0^2-x_3^2-(x_1^2+x_2^2-l^2)$ which will occur. Therefore, we
shall have a again the locality condition in the form:
$$
[j_1(\frac{x}{2}),j_2(-\frac{x}{2})]=0\ , \ \mbox{for}\ \
x_0^2-x_3^2-(x_1^2+x_2^2-l^2)< 0\ ,
$$
which is equivalent to (\ref{newcaus}).

The new causality condition (\ref{newcaus}) is strongly supported
by calculations performed in the first paper dealing with the
causality in NC QFT, in Lagrangian approach, \cite{causal1}. There
it was shown, through the study of a scattering process, that
space-space NC $\phi^4$ in 2+1 dimensions is causal at
macroscopical level\footnote{We would like to emphasize that, if
one admits the causality condition in the form (\ref{caus}) for
3+1-dimensional NC QFT, then for the 2+1-dimensional theory
treated in \cite{causal1} one simply could not write any
(micro)causality condition, since all (two) spacial coordinates
are noncommutative and the signal should propagate instantaneously
in all directions.}. Moreover, the same study \cite{causal1} has
shown that the scattered wave appears to originate from a position
shifted by $\frac{1}{2}\theta p$, where $p$ is the momentum of the
incoming wave packet. The physical interpretation given in
\cite{causal1} is that the incident particles should be viewed as
extended rigid rods, of the size $\theta p$, perpendicular to
their momentum. In other words, the noncommutativity introduces a
scale $\theta$ of the spacial nonlocality. The effect calculated
in \cite{causal1} is actually an amplification at macroscopic
scale of the (micro)causality condition (\ref{newcaus}).

Correspondingly, the spectral condition will read as
\be\label{newspectr} p_0^2-p_1^2-p_2^2-p_3^2\geq 0,\ \ \ p_0>0\ .\ee
This is the case since a NC QFT with $\theta_{\mu\nu}$ a constant
matrix has the twisted \P\ symmetry. The latter, however, has the
representation content identical to the usual \P\ symmetry and thus
the particles dispersion law and the spin structure are the same as
the usual (commutative) QFT case  \cite{CKT}. Thus both the
causality condition (\ref{newcaus}) and the spectral condition
(\ref{newspectr}) are given in terms of the invariants of the
theory.

In fact, the consideration of nonlocal theories of the type
(\ref{newcaus}) was initiated by Wightman \cite{nonloc}, who asked
the concrete question whether the vanishing of the commutator of
fields (or observables), i.e.
$[j_1(\frac{x}{2}),j_2(-\frac{x}{2})]=0$, for
$x_0^2-x_1^2-x_2^2-x_3^2<-l^2 $ would imply its vanishing for
$x_0^2-x_1^2-x_2^2-x_3^2<0$. It was proven later \cite{Vlad,
petrina, Wight} (see also \cite{BLT}) that, indeed, in a quantum
field theory which satisfies the translational invariance and the
spectral axiom (\ref{newspectr}), the nonlocal commutativity
$$
[j_1(\frac{x}{2}),j_2(-\frac{x}{2})]=0\ ,\ \ \ \mbox{for}\ \ \
x_0^2-x_1^2-x_2^2-x_3^2<-l^2
$$
implies the local commutativity
\be\label{comm} [j_1(\frac{x}{2}),j_2(-\frac{x}{2})]=0\ ,\ \ \
\mbox{for}\ \ \ x_0^2-x_1^2-x_2^2-x_3^2<0\ . \ee

This powerful theorem, which does not require Lorentz invariance,
can be applied in the \nc case, since the hypotheses are
fulfilled, with the conclusion that the causality properties of a
QFT with space-space \ncy are physically identical to those of the
corresponding commutative QFT.

It is then obvious that the Jost-Lehmann-Dyson representation
(\ref{c_jld}) obtained in the commutative case holds also on the
NC space. Consequently, the NC two-particle$\to$two-particle
scattering amplitude will have the same form as in the commutative
case:
\be M(E,\cos\Theta)=i\int d^4 u d\kappa^2\frac{\phi( u, \kappa^2,
k+p)} {\left[\frac{1}{2}( k'- p')+ u\right]^2-\kappa^2}\ .\ee
This leads to the analyticity of the NC scattering amplitude in
$\cos\Theta$ in the analog of the Lehmann ellipse, which behaves
at high energies $E$ the same way as in the commutative case, i.e.
with the semi-major axis as
\be y_L=(\cos\Theta)_{max}=1 +\frac{const}{E^4}\ . \ee

\section {Enlargement of the domain of analyticity in $\cos\Theta$
and use of unitarity. Martin's ellipse}

Two more ingredients are needed in order to enlarge the domain of
analyticity in $\cos\Theta$ to the Martin's ellipse and to obtain
the Froissart-Martin bound: the dispersion relations and the
unitarity constraint on the partial-wave amplitudes \cite{review}.

We recall the conclusion of \cite{Liao} that, when using the
causality condition (\ref{causal}), the forward dispersion
relation cannot be obtained in NC theory with general direction of
the $\vec\beta$-vector. However, the conclusion to which we
arrived by imposing a physical nonlocal commutativity condition
(\ref{newcaus}) and reducing it to the local commutativity
(\ref{comm}), by using the theorem due to Wightman, Vladimirov and
Petrina, leads straightforwardly to the usual forward dispersion
relation also in the NC case with a general $\vec\beta$ direction.
We have to point out that now the derivations of both the Lehmann
ellipse and of the forward dispersion relation do {\it not} depend
on the orientation of the incoming momentum $\vec p$ with respect
to the NC plane, or equivalently to the $\vec\beta$-vector.

As for the unitarity constraint on the partial wave amplitudes,
the problem has been dealt with in \cite{CMT}, for a general case
of noncommutativity $\theta_{\mu\nu}$, $\theta_{0i}\neq 0$. For
space-space noncommutativity ($\theta_{0i}= 0$), the scattering
amplitude depends, besides the center-of-mass energy, $E$, on
three angular variables. In a system were we take the incoming
momentum $\vec p$ in the $z$-direction, these variables are the
polar angles of the outgoing particle momentum, $\Theta$ and
$\phi$, and the angle $\alpha$ between the vector $\vec\beta$ and
the incoming momentum. The partial-wave expansion in this case
reads:
\be\label{pwe} A(E,
\Theta,\phi,\alpha)=\sum\limits_{l,l',m}(2l'+1)a_{ll'm}(E)Y_{lm}(\Theta,\phi)P_{l'}(\cos\alpha)\
, \ee
where $Y_{lm}$ are the spherical harmonics and $P_{l'}$ are the
Legendre polynomials.

Imposing the unitarity condition directly on (\ref{pwe}) or using
the general formulas given in \cite{CMT}, it can be shown that a
simple unitarity constraint which involves single partial-wave
amplitudes one at a time cannot be obtained in general, but only
in a setting where the incoming momentum is orthogonal to the NC
plane (equivalently it is parallel to the vector $\vec\beta$). In
this case the amplitude depends only on one angle, $\Theta$, and
the unitarity constraint is reduced to the well-known one of the
commutative case, i.e.
\be Im\ a_l(E)\geq |a_l(E)|^2\ . \ee

For this particular setting, $\vec p \parallel\vec \beta$, it is
then straightforward, following the prescription developed for
commutative QFT, to enlarge the analyticity domain of scattering
amplitude to Martin's ellipse with the semi-major axis at high
energies as \be y_M=1 +\frac{const}{E^2}\ \ee and subsequently
obtain the NC analog of the Froissart-Martin bound on the total
cross-section, in the CMS and for $\vec p\parallel\vec \beta$:
\be\label{froissart} \sigma_{tot}(E)\leq c\ \ \ln^2 \frac{E}{E_0}\
. \ee

In the same manner, rigorous high-energy bounds on the nonforward
scattering amplitude,
\be\label{nonforward} |A(E, \cos\Theta)|<const\
E^{3/2}\ln^{3/2}\left(\frac{E}{E_0}\right)\frac{1}{|\sin\Theta|^{1/2}}\
,\ee
as well as on the differential cross-section,
\be\label{diff}
\left(\frac{d\sigma}{d\Omega}\right)|_{\Theta=0}<const\ E^2
\ln^{4}\frac{E}{E_0}\ ,\ee
are obtained in the NC case with $\vec p \parallel\vec \beta$. The
high-energy behaviour of these bounds is identical to the one in
the ordinary QFT.

Thus, the unitarity constraint on the partial-wave amplitudes
distinguishes a particular setting ($\vec p\parallel\vec \beta$)
in which the Lehmann's ellipse can be enlarged to the Martin's
ellipse and Froissart-Martin bound can be obtained. However, since
the Lehmann's ellipse {\it is} the domain of analyticity in
$\cos\Theta$ irrespective of the direction of $\vec p$, one might
attempt to obtain at least a NC analog of the Greenberg-Low bound
(\ref{gl}) for $\vec p \nparallel\vec\beta$. However, according to
the orthodox procedure for the derivation of this bound \cite{GL},
besides analyticity in Lehmann's ellipse and polynomial
boundedness, a suitable unitarity constraint on the partial-wave
amplitudes is still needed, which we have not succeeded to derive.
Nevertheless, this does not exclude the possibility of obtaining a
rigorous high-energy bound on the cross-section for $\vec p
\nparallel\vec\beta$, and the issue deserves further
investigation.

\section{Generalization to the flat (noncompact)\\ higher-dimensional NC space-time}

In this section we shall derive rigorous high-energy bounds on the
total cross-section, on the two-particle$\to$two-particle
nonforward scattering amplitude and on the differential
cross-section for the case of NC space-time with an arbitrary
number of {\it noncompact} dimensions $D$. For the usual
(commutative) case of higher-dimensional space-time, such
high-energy bounds have been previously derived
\cite{Fischer,CFV}.

Due to the recent activity in the NC version of higher-dimensional
theories (see, e.g., \cite{highdim} and references therein), the
generalization of high-energy bounds derived in the previous
section to higher dimensions is of interest, especially since in
this case there appear distinct features between the theories in
which the space-time dimension $D$ is odd or even. Specifically,
for the case of even $D$ we shall be able to derive high-energy
bounds, analogous to the bounds in usual commutative QFT, for the
special setting when the incoming particle momentum is in the
direction of the 'magnetic' vector $\vec\beta$ (to be defined
below).

Consider QFT on NC space-time of arbitrary dimension $D$, with
\be\label{crhigh} [x_\mu,x_\nu]=i\theta_{\mu\nu}\ ,\ \ \
\mu,\nu=0,1,...,D-1\ \ee
and $\theta_{\mu\nu}=-\theta_{\nu\mu}$ arbitrary real constants.
To avoid the known problems in the case $D=4$ with unitarity
\cite{unit} and causality \cite{causal1,causal2}, which will
appear automatically in the general case of $D$ dimensions, we
should have that the noncommutativity is only between spacial
components, i.e. $\theta_{0i}=0$, $i=1,...,D-1$ (or equivalently
there is no 'electric' vector\footnote{For terminology, see
footnote on the definition of vector $\vec\beta$ in Section 3.}
$\vec\epsilon$, $\epsilon_i\equiv\theta_{0i}$.)

We also mention a crucial point that for a
two-particle$\to$two-particle scattering, the amplitude still
depends on only two kinematical variables, $E$ and $\cos\Theta$,
for any dimension $D$ of ordinary (commutative) space-time
\cite{Fischer}. In this case the symmetry is $SO(1,D-1)$ and the
partial-wave expansion is in terms of the Gegenbauer (instead of
Legendre) polynomials, for any $D$, and there is no distinction in
the final results between odd and even $D$.

Consider now the case of a space-time of even dimension $D$, with
commutation relation (\ref{crhigh}) and $\theta_{0i}=0$. Any
$D\times D$ antisymmetric matrix $\theta_{\mu\nu}$ possesses
paired eigenvalues $\pm\lambda_{\alpha}$, with
$\alpha=1,...,\frac{D}{2}$. Due to the conditions $\theta_{0i}=0$,
$i=1,...,D-1$, there is one pair of light-like vectors
corresponding to $\lambda=0$. We choose this pair of vectors to
span the plane $(0,D-1)$. With this choice, the 'magnetic' vector
$\vec\beta$, defined as
\be\label{betahigh}
\beta_i=\epsilon_{ii_1i_2...i_{D-2}}\theta_{i_1i_2}...\theta_{i_{D-3}i_{D-2}}
\ee
becomes oriented to the $(D-1)$-direction, i.e.
$\beta_1=\beta_2=...=\beta_{D-2}=0$ and $\beta_{D-1}=det\theta'$
is the determinant of the $(D-2)\times(D-2)$ matrix
$\theta_{i'j'}$, with $i',j'=1,...,D-2$. We shall consider the
case $det \theta'\neq 0$, i.e. when there are no more zero
eigenvalues $\lambda_{\alpha}$. For the generic case with
distinct, nonzero eigenvalues of $\theta_{i'j'}$, the space-time
symmetry is $SO(1,1)\times\prod\limits_{l=1}^{(D-2)/2}SO_{(l)}(2)$
(when the eigenvalues are degenerate, the symmetry is enlarged).

For the general relative direction of incoming particle momentum
$\vec p$ and the 'magnetic' vector $\vec\beta$, a partial-wave
expansion of the amplitude can be written down analogous  to the
one given for $D=4$ in \cite{CMT}. However, for the special
configuration where the incoming particle momentum $\vec p$ is
parallel to $\vec\beta$, the number of kinematical variables
entering the scattering amplitude becomes two, $E$ and
$\cos\Theta$, and in this case the partial-wave expansion of
$D$-even dimensional scattering amplitude is analogous to the one
treated in \cite{Fischer,CFV}.

With the same physical arguments on the finiteness of the
instantaneous nonlocal interaction in the NC hyperplane, as
presented in Section 3 and leading to the causality condition
(\ref{newcaus}), we shall have now the causality condition as:
\be\label{lochigh} [j_1(\frac{x}{2}),j_2(-\frac{x}{2})]=0\ ,\
\mbox{for}\ x_0^2-x_{D-1}^2-(x_1^2+...+x_{D-2}^2-l^2)<0\ .\ee
Using now the extension of the Wightman-Vladimirov-Petrina theorem
\cite{Vlad,petrina,Wight} to an arbitrary dimension $D$, one
arrives to the conclusion that the nonlocality condition of the
type (\ref{lochigh}) implies the ordinary causality condition:
\be\label{caushigh} [j_1(\frac{x}{2}),j_2(-\frac{x}{2})]=0\ ,\
\mbox{for}\ x_0^2-x_{D-1}^2-(x_1^2+...+x_{D-2}^2)<0\ .\ee

As a final result, for an even-$D$-dimensional NC space-time, for
the special kinematical configuration of $\vec
p\parallel\vec\beta$, we obtain the rigorous high-energy bounds on
the total cross-section, nonforward scattering amplitude and on
the differential cross-section:
\bn \label{b1}\sigma_{tot}(E)&<& c\ \ \left(\ln\frac{E}{E_0}\right)^{D-2}\ ,\\
\label{b2}|A(E, \cos\Theta)|&<&const\
E^{(7-D)/2}\left(\ln\frac{E}{E_0}\right)^{(D-1)/2}\frac{1}{|\sin\Theta|^{(D-3)/2}}\ ,\\
\label{b3}\left(\frac{d\sigma}{d\Omega}\right)|_{\Theta=0}&<&const\
E^2 \left(\ln\frac{E}{E_0}\right)^{2(D-2)}\ .
 \en

The bounds (\ref{b1}), (\ref{b2}) and (\ref{b3}) are the analogs
of the Froissart and other bounds obtained previously for the
ordinary (commutative) QFT in $D$ dimensions \cite{Fischer,CFV},
identical in their high-energy behaviour.

We mention that for the case of an odd dimension $D$ of
space-time, one cannot construct a 'magnetic' vector $\vec\beta$
such as (\ref{betahigh}). Instead, there exists a 2-dimensional
plane, orthogonal to the $(D-3)$-dimensional NC hyperplane and
thus no kinematical configuration for the incoming particle
momentum is possible, in order that no extra angular variables
would appear. As a result, no simple unitarity constraint, which
would involve single partial-wave amplitudes one at a time, can be
derived. Thus, although there exists the Lehmann ellipse for the
analyticity domain in $\cos\Theta$  of the
two-particle$\to$two-particle scattering amplitude, no extension
to Martin's ellipse is possible. Consequently, not only the
derivation of a Froissart-Martin bound in $D$ odd NC case is
impossible, there seems no way to rigorously obtain even a weaker
bound analogous to the Greenberg-Low bound.

\section{Conclusion and discussions}

In this paper we have tackled the problem of high energy bounds on
the two-particle$\to$two-particle scattering amplitude in NC QFT
with $SO(1,1)\times SO(2)$ symmetry. We have found that, using the
causal and spectral conditions (\ref{causal}) and (\ref{spectr})
proposed in \cite{LAG} for NC theories with such symmetry, a new
form of the Jost-Lehmann-Dyson representation (\ref{4.32}) is
obtained, which does not permit to draw any conclusion about the
analyticity of the scattering amplitude (\ref{scat}) in
$\cos\Theta$. However, the physical observation that nonlocality in
the noncommuting coordinates is not infinite brought us to imposing
a new causality condition (\ref{newcaus}), which accounts for the
finitness of the range of nonlocality and prevents the {\it
instantaneous propagation of signals} in the {\it entire}
noncommutative plane $(x_1,x_2)$. We proved that the new causality
condition is formally identical to the one corresponding to the
commutative case (\ref{comm}), using the Wightman-Vladimirov-Petrina
theorem \cite{Vlad,petrina,Wight}. We would like to mention that
there the NC QFT with noncommutativity given by (\ref{cr}) possesses
the twisted Poincar\'e symmetry which preserves the commutation
relation (\ref{cr}) between the coordinates, while the
representation content and the invariants are identical to the ones
of ordinary Poincar\'e algebra \cite{CKT}. In this sense, the
causality and the spectral conditions in the forms (\ref{newcaus})
and (\ref{newspectr}) are described by invariants of the theory, as
it should be.

Thus, the scattering amplitude in NC QFT is proved to be
analytical in $\cos\Theta$ in the Lehmann ellipse, just as in the
commutative case; moreover, dispersion relations can be written on
the same basis as in commutative QFT.

Finally, based on the unitarity constraint on the partial-wave
amplitudes in NC QFT, we can conclude that, for theories with
space-space noncommutativity ($\theta_{0i}=0$), the total
cross-section is subject to an upper bound (\ref{froissart})
identical to the Froissart-Martin bound in its high-energy
behaviour, when the incoming particle momentum $\vec p$ is
orthogonal to the NC plane. In the same configuration, similar
bounds on nonforward scattering amplitude (\ref{nonforward}) and
on the differential cross-section (\ref{diff}) are obtained. This
is the first example of a {\it nonlocal} theory, in which
cross-sections do have an upper high-energy bound.

Furthermore, our considerations have been generalized to the case of
NC space-time of an arbitrary dimension $D$. For the case of even
dimension D, in the specific configuration in which the incoming
particle momentum $\vec p$ is parallel to the 'magnetic' vector
$\vec\beta$, the generalization of the bounds on the cross-sections
(\ref{b1}), the nonforward scattering amplitude (\ref{b2}) and on
the differential cross-section (\ref{b3}) are obtained. In the case
of an odd dimension $D$ of NC space-time, although there exists the
Lehmann ellipse as analyticity domain of the scattering amplitude in
$\cos\Theta$, there seems to be no way to derive even weaker bounds,
such as the Greenberg-Low bound \cite{GL}. This is due to the
appearance of additional angular variables, thus leading to a
complicated form of the unitarity constraint on the partial-wave
amplitudes, which seemingly makes its utilization impossible, at
least in its present form.

There remain several interesting questions to be studied. Among
them is the question whether any bound can in principle be
obtained for the case when the incoming particle momentum is not
orthogonal to the NC plane, or, equivalently, is not parallel to
the 'magnetic' vector. Similar questions appear also in the case
of NC QFT in a space-time of arbitrary dimension $D$, concerning
the distinct features between even and odd $D$, as considered in
Section 5 of the paper.

\vskip 0.3cm {\Large\bf{Acknowledgements}}

We are grateful to L. \'Alvarez-Gaum\'e, C. Montonen, H. Miyazawa
and K. Nishijima for useful discussions. We are indebted to P.
Pre\v{s}najder for his enlightening remarks and suggestions. We
thank M. N. Mnatsakanova and Yu. S. Vernov for their collaboration
at an early stage of this work.

The financial support of the Academy of Finland under the Projects
No. 54023 and 104368 is acknowledged.

\end{document}